# Long-range Quantum Cryptography: Amplified Quantum Key Distribution (AQKD)*

Richard J. Hughes and Jane E. Nordholt
Los Alamos National Laboratory
Los Alamos, NM 87544, USA

The principal remaining technical impediments to the wide adoption of quantum key distribution (QKD) in its present form are: its limitation to metro-area fiber spans of typically less than 100km (~ 200km with heroic measures); and its incompatibility with the optical amplifiers (OA) that are typically installed with ~ 70-km spacing on longer fiber spans. Trusted QKD networks [1] and quantum repeater networks can overcome the metro-area range limit, but are not compatible with OAs and so cannot be deployed as overlays on existing fiber infrastructure. Further, the dedicated network resources, requirements for multiple intermediate nodes, and duplication of quantum resources make these architectures cumbersome, expensive and hence of limited practical potential. We present a new protocol called amplified quantum key distribution (AQKD) [2] that is compatible with OAs, and describe a simulation of a particular AQKD configuration that would provide as much as twice the single-span range of standard weak-laser BB84 QKD. With AQKD, inter-city distances such as Boston-New York will become possible without the practical drawbacks of the trusted relay and quantum repeater approaches.

QKD over amplified spans is considered infeasible because the amplified spontaneous emission (ASE) [3] noise added by an OA would push the sifted bit error rate (BER) above the upper limit of the standard BB84 protocol. However, for a classical broadcast channel Maurer [4] has shown that Alice and Bob can use a protocol called advantage distillation (AD) to achieve a non-zero secret key capacity, even if Bob's signal-to-noise ratio (SNR) is initially worse than Eve's. A first condition for using AD is that Alice and Bob know a rigorous lower bound on Eve's SNR, which is problematic for classical communications. But for quantum communications, if Alice's transmitter includes an OA it will introduce unavoidable noise of quantum origin (due to ASE) that will rigorously enforce such a bound on Eve's SNR by the laws of quantum physics. A second condition for using AD is that Bob's and Eve's errors are uncorrelated, which will hold if Eve is restricted to passive tapping of the Alice-Bob quantum channel. For this case, the recently-established loose upper bound on secret key capacity (referred to here as the "Takeoka bound") [5] is more than an order-of-magnitude higher than the capacity of weak-laser BB84 QKD for typical channel parameters. Together, these observations suggest that major performance gains are achievable under this security model using our new AQKD protocol (Fig. 1), in which the Alice-Bob quantum channel contains one or more OAs, and the protocol post-processing stage includes AD after sifting, but before error correction (EC) and privacy amplification (PA) against tapping.

| Transaction | Content | Result |
|---|---|---|
| A→B | qubits | raw key |
| B→A | advantage distillation | |
| A→B | advantage distillation | distilled key |
| A→B | EC information | reconciled key |
| A→B | PA function | final key |
| A→B | authentication tag | |
| B→A | authentication tag | |

Figure 1. Amplified quantum key distribution (AQKD) protocol between Alice (A) and Bob (B). See text for details.

We demonstrate these gains with a simulation of the polarization-coded AQKD instantiation of Fig. 2, in which Alice's qubit states (Poisson-distributed photon number with mean $\mu \sim 1$ at ~ 1,550-nm wavelength), produced by our integrated photonics BB84 "QKarD" light source [6], are amplified by a phase-insensitive OA. Alice's OA, with gain $G$, produces ASE in each polarization mode with mean-photon number $\langle n_{sp} \rangle = \chi(G - 1)$ per longitudinal mode [7], where $\chi \geq 1$ is the excess noise factor. In the following we assume $\chi = 1$, the well-known "3-dB limit" of unavoidable, minimum noise of quantum origin of a phase-insensitive amplifier. After passing through an $L$-km span of single-mode fiber (SMF) these signals are detected by Bob's passive-basis choice BB84 receiver. We assume polarization errors of

*Research funded by DARPA Quiness program

1%, and single-photon detector (SPD) characteristics (detection efficiency, $\eta_d$, and dark noise probability per gate, $p_d$) that are representative of ns-gated indium-gallium arsenide (InGaAs) avalanche photodiodes (APD) operated in Geiger mode. Alice's and Bob's enclaves also include matched optical filters, whose bandwidth can be selected to achieve single longitudinal mode operation. (Filters of ~ 1-GHz, 3-dB bandwidth, which would be required for 1-ns detector gates, are commercially available). With 30dB out-of-band rejection, these filters also suppress out-of-band ASE and fiber-generated Raman-noise [8] to negligible levels. A suitable polarization tracking and compensation system to correct fiber birefringence has been demonstrated [6], but is not shown in Fig. 2.

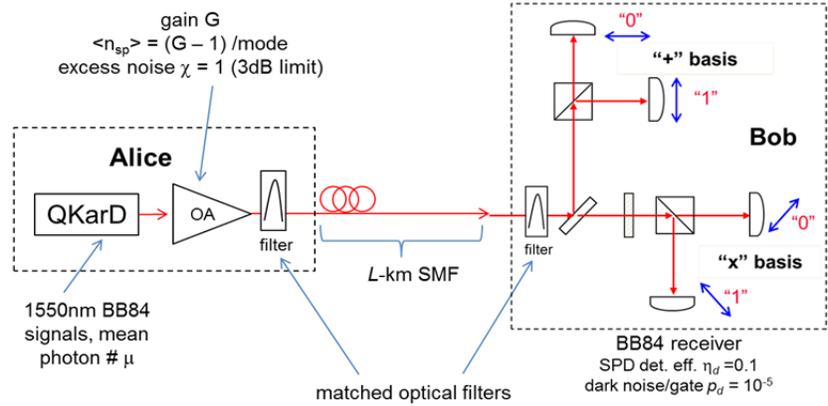

Figure 2. A representative AQKD configuration in which Alice's BB84 signals are optically amplified before transmission to Bob. See text for details.

Alice's amplified BB84 signals contain a Laguerre-Gauss distributed photon number of mean $n_{LG} = G\mu + <n_{sp}>$ in the same polarization, and a Bose-Einstein distributed photon number with mean $n_{BE} = <n_{sp}>$ in the orthogonal polarization, resulting in quantum signals of mean photon number $n = (n_{LG} + n_{BE})$ and polarization fidelity $F = n_{LG}/n$. For $G > 1$ Bob's sifted key therefore has more bits and a higher BER than in standard BB84 ($G = 1$) at the same $\mu$ value. In the passive tapping scenario we assume that Eve receives *every* photon that Bob does not detect, and uses a passive-basis choice BB84 receiver with ideal (100% efficient, zero noise) photon-number resolving detectors. By monitoring Alice's and Bob's sifting messages, Eve learns the sifting basis for each bit. Within that basis, she assigns an unambiguous sifted bit value to whichever of her two detectors records the larger photo-count, and a random bit value for the ambiguous case when the two detectors record equal photo-counts. Neither Eve's unambiguous bit errors nor her ambiguous bits are correlated with Bob's sifted bit errors. In the next, distillation, phase of the AQKD protocol we both generalize Maurer's AD to accommodate both unambiguous and ambiguous portions of Eve's sifted key, and simplify its implementation.

To form their distilled keys Alice, Bob (and Eve) parse their $N$-bit sifted keys $\{x_j, j = 0, 1, \ldots N - 1\}$, where $x_j = a_j$ for Alice (A), $x_j = b_j$ for Bob (B), and $x_j = e_j$ for Eve (E) respectively, into $N/2$ bit-pairs $[x_{2i+1}, x_{2i}]$, $i = 0, 1, \ldots (N/2 - 1)$. (We assume $N$ is even without loss of generality.) Next, Alice, Bob and Eve each form two, $N/2$-bit subsequences: $F = \{f_i = x_{2i}, i = 0, 1, \ldots (N/2 - 1)\}$, and $P = \{p_i = x_{2i+1} \oplus x_{2i}, i = 0, 1, \ldots (N/2 - 1)\}$. Alice transmits her parity sequence, $P_A = \{a_{2i+1} \oplus a_{2i}, i = 0, 1, \ldots (N/2 - 1)\}$, to Bob. He compares the value $p_{A,i}$ of each element of the sequence $P_A$ received from Alice, with the corresponding element $p_{B,i}$ of his sequence $P_B$. From his bit sequence $F_B$ he forms his distilled key, $D_B$, of bits $f_{B,i}$ for which $p_{B,i} = p_{A,i}$, i.e. $D_B = \{f_{B,i} \mid p_{B,i} = p_{A,i}, i = 0, 1, \ldots (N/2 - 1)\}$, and discards the rejected bits of $F_B$ (i.e. bits $f_{b,i}$ for which $p_{B,i} \neq p_{A,i}$) as well as the entire parity sequence $P_B$. Bob calculates the rejection index sequence, $R = \{i \mid p_{B,i} \neq p_{A,i}, i = 0, 1, \ldots (N/2 - 1)\}$, and sends it to Alice. With this information, Alice discards the rejected bits from her sequence $F_A$, as well as her entire sequence $P_A$, to form her distilled key, $D_A$. To form her distilled key, $D_E$, Eve can assign an unambiguous bit value to each distilled bit arising from a pair of unambiguous sifted bits, but must assign a random (ambiguous) value to all other distilled bits. This generalized advantage distillation (GAD) protocol results in distilled keys on which Bob's BER is lower than on his sifted key, while Eve's ambiguous fraction is increased and the BER on her unambiguous portion is unchanged. Alice and Bob now form their secret keys by applying EC and PA against passive tapping to their distilled keys. Although the reduction in key size for GAD somewhat offsets the increase from optical gain, Eve's reduced information per bit (owing to ASE and distillation) results in more secret key bits for Alice and Bob than in standard BB84 at the same $\mu$.

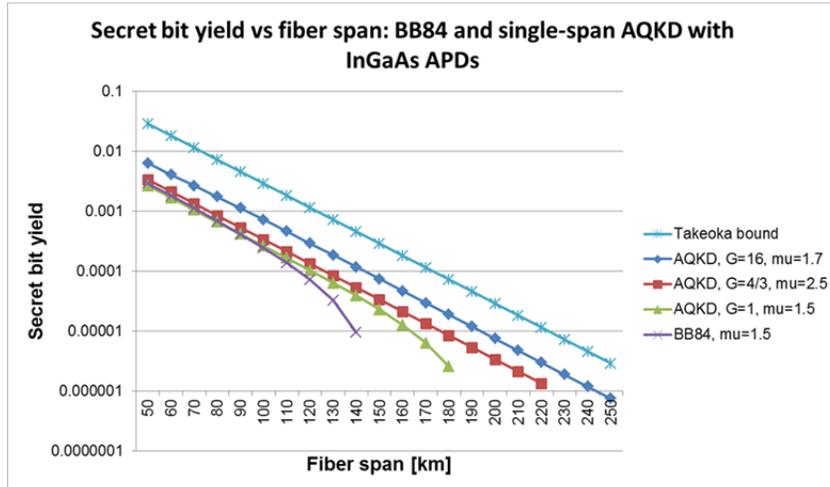

Figure 3. Simulation results for the secret bit yield (secret bits per transmitted bit) as a function of fiber span length for the system of Fig. 1 under the passive tapping security model. See text for details.

This is apparent in Fig. 3, where the lowest, purple curve shows the secret bit yield of standard BB84 at the optimal mean photon number of $\mu = 1.5$ under the passive tapping security model. Up to 100-km span lengths, standard BB84 is an order of magnitude below the Takeoka bound (upper-most, light-blue curve). Taking a practical lower bound of $10^{-6}$ on secret bit yield as defining the maximum range, we find that standard BB84 cannot go beyond 145km in SMF, whereas the Takeoka bound is ~ 275km. For the green curve, we have introduced a stage of GAD after sifting and before EC into standard BB84, without optical gain ($G = 1$). We see that this gives no improvement over standard BB84 for spans up to 100km, but does increase the maximum range by 40km. This illustrates how GAD is particularly effective in low SNR regimes. With the red curve, we introduce a gain of $G = 4/3$ (corresponding to optimal 1→2 quantum cloning), and find optimal AQKD performance for a mean photon number input to Alice's OA of $\mu = 2.5$. Even at this low gain AQKD gives some improvement in secret bit yield over standard BB84 for spans up to 100km (~10% increase), while increasing the maximum range by 50% to 210km. The dark blue curve is for a larger gain of $G = 16$, with a mean photon number input to Alice's OA of $\mu = 1.7$. For this configuration, AQKD gives: ~ 3 times the secret bit yield of standard BB84 at span lengths up to 100km; more than an order of magnitude improvement at 140km; and adds more than 100km to the maximum range of standard BB84 (to 250km), while coming within a factor of 4 of the Takeoka bound. The maximum range for this configuration could be further pushed out to ~ 300km if Bob uses high-efficiency, low-noise superconducting detectors, and ultra-low loss SMF is used for the Alice-Bob fiber. We expect that performance gains (range and secret rate) will also result from adding one or more intermediate OAs between Alice and Bob to the single-OA configuration of Fig. 1. An experimental implementation of Fig.1 to evaluate AQKD performance with practical OAs is now in preparation for our QKD network test bed [6] at Los Alamos. We are investigating the extension of the AD and PA stages of the AQKD protocol for security against active attacks (intercept-resend, photon-number splitting etc.): such attacks will give Eve less information than in standard BB84, owing to the ASE noise.

Our AQKD protocol will extend the range of quantum cryptography from metro-area to at least inter-city distances, and we expect that it can be deployed as an overlay on fiber spans carrying conventional optical traffic [6, 8] that contain intermediate OAs. It will enable the wide application of QKD.